\documentclass{article}
\usepackage{spconf,amsmath,graphicx,xcolor,hyperref}
\hypersetup{hidelinks}
\usepackage{booktabs}
\usepackage{makecell}
\usepackage{threeparttable}
\usepackage{amssymb}
\usepackage{enumitem}
\usepackage{cite}
\usepackage{stfloats}
\usepackage[numbers,sort&compress]{natbib} 
\usepackage{caption}
\newcommand{\paperauthor}[3]{%
  \textit{#1}$^{#2}$%
}

\newcommand{\Huang}[1]{{\color{black}{#1}}}

\title{Curv-Tail: Lightweight Long-Tailed Encrypted Traffic Classification with Discrete Packet-Length Encoding and Lorentz Prototypes}

\name{%
  \begin{tabular}{@{}c@{}}
    \paperauthor{Yankun Wang}{1}{},
    \paperauthor{Jun-Jie Huang}{1}{},
    \paperauthor{Lin Liu}{1}{},
    \paperauthor{Xiaodong Lei}{1}{},
    \paperauthor{Yi Chen}{1}{} \\[-0.25ex]
    \paperauthor{Yeqing Yan}{2}{},
    \paperauthor{Lin Liu}{1}{},
    \paperauthor{Jiangyong Shi}{1}{},
    \paperauthor{Yongjun Wang}{1}{}
  \end{tabular}%
 \thanks{\fontsize{9}{11}\selectfont
The third and seventh authors, both named Lin Liu,
are different individuals.}
}
\address{%
  $^{1}$National University of Defense Technology, Changsha, China\\
  $^{2}$Hunan First Normal University, Changsha, China
}

\begin{document}
%
\maketitle
\begin{abstract}
Long-tailed encrypted traffic classification requires accurate recognition of
infrequent classes under limited computational budgets. We propose Curv-Tail,
a lightweight, end-to-end packet--byte framework trained without a separate
pretraining stage. Mixed-resolution tokenization preserves exact packet-length
identities within a bounded range and coarsens larger values to limit the
vocabulary. An auxiliary objective predicts observed length tokens from
contextual packet features before pooling, encouraging length-token retention
beyond flow-level supervision. Compact temporal encoders process packet
sequences and directional byte patches, and Lorentz prototypes with a shared
learnable curvature magnitude classify their fused representation.
On NUDT-Mobile and DataCon-Website under natural class frequencies, Curv-Tail
achieves three-seed mean Tail-F1 scores of 85.70\% and 46.30\%, exceeding the
strongest evaluated baselines by 2.91 and 1.89 percentage points, respectively.
In 300-class profiling on an RTX 4090 with FP32 and batch size 256, Curv-Tail
uses 98.48\% fewer parameters and achieves 10.2$\times$ the batch inference
throughput of MM4Flow.
\end{abstract}

\begin{keywords}
Encrypted traffic classification, long-tailed classification, packet-length
representation, Lorentz geometry, lightweight models.
\end{keywords}
\section{Introduction}
\label{sec:intro}

Encrypted traffic classification identifies applications and websites from
observable traffic features without decrypting payloads, supporting network
monitoring and application-aware traffic management. Traffic collected in
practice often exhibits long-tailed class frequencies: a few popular services
generate abundant flows, whereas many infrequent services 

contribute relatively few samples~\cite{wickramasinghe2025sok,hu2026lotbonc}.
Limited labeled flows make it difficult to capture within-class variability for tail classes.
Because overall accuracy weights classes by their test-flow counts, strong head-class performance can mask errors on infrequent classes. 
Tail-class recognition alongside aggregate performance is therefore important
for this setting.

Existing classifiers exploit packet lengths, byte representations, and multiple
traffic views~\cite{Shen2021GraphDapp,lin2022bert,zhao2023yet,
yang2025mm4flow,zhou2025trafficformer}.
For long-tailed recognition, we evaluate classifier-level corrections on ET-BERT~\cite{lin2022bert} after fine-tuning and freezing its encoder, following a decoupled learning setup~\cite{kang2019decoupling}. On NUDT-Mobile~\cite{ZhaoShuang24NudtAPP}, the best evaluated correction improves Tail-F1 by only 0.08 percentage points (Table~\ref{tab:longtail_comparison_etbert_nudt}). This limited gain motivates examining representation learning beyond the evaluated corrections. We pursue this direction with a compact packet--byte framework that jointly learns from both views while controlling storage and inference costs.

\begin{figure}[t]
    \centering
    \includegraphics[width=0.85\columnwidth]{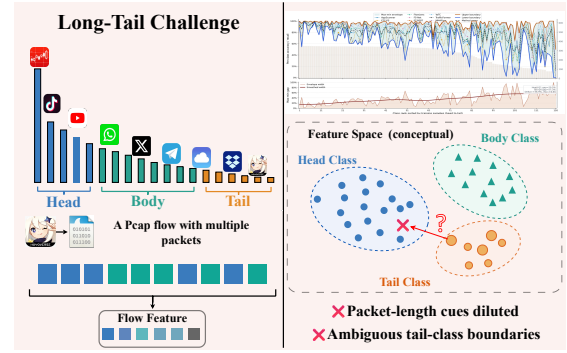}
\caption{Long-tailed encrypted traffic classification and its two main challenges.}
    \label{fig:Challenge}
\end{figure}

Fig.~\ref{fig:Challenge} motivates two design goals: preserving packet-length cues and improving tail-class separation. First, numerical proximity between lengths need not imply similar discriminative value, and flow-level supervision does not explicitly require contextual packet features to retain length information before pooling. These considerations motivate discrete length tokens and an auxiliary retention objective within the packet branch. Second, scarce tail-class samples complicate learning reliable class boundaries, motivating us to investigate prototype-based classification of the fused packet--byte representation.

\begin{figure*}[t]
    \centering
    \includegraphics[width=0.95\textwidth]{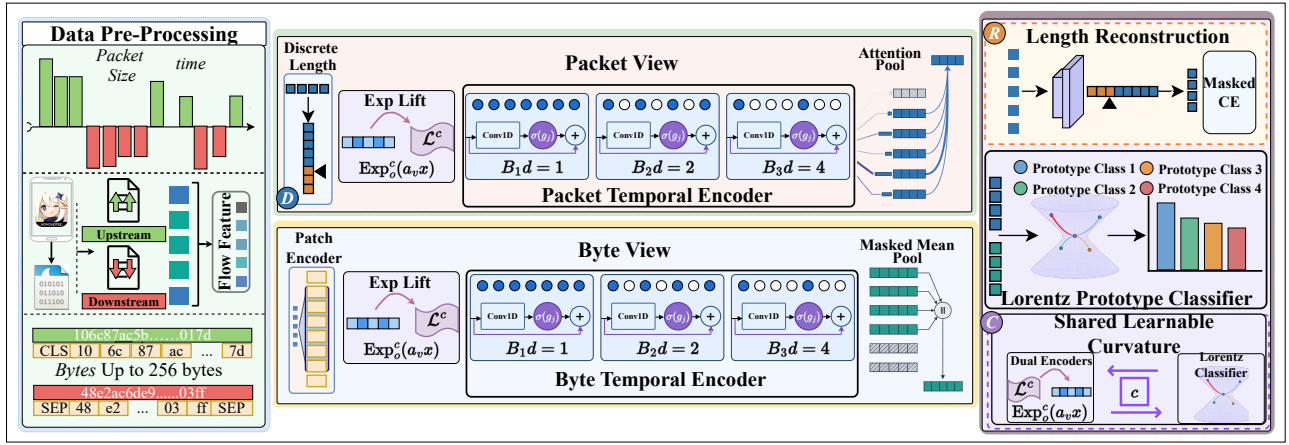}
    \caption{Overview of Curv-Tail.
    }
    \label{fig:curv_tail_architecture}
    \vspace*{-0.6cm}
\end{figure*}

We propose Curv-Tail\textsuperscript{\ref{fn:curvtail-code}}, a lightweight, end-to-end framework for long-tailed
encrypted traffic classification.

Mixed-resolution tokenization preserves exact length identities within a bounded range, and an auxiliary objective encourages their retention in contextual packet features before pooling.
Two compact temporal branches encode packet sequences and directional byte patches.
Building on prototype classification~\cite{ijcai2023p151}, hyperbolic prototype learning~\cite{NEURIPS2021_01259a0c}, and curvature learning~\cite{fan2025curvature}, we use Lorentz prototypes with a shared learnable curvature magnitude to classify the fused representation.
The framework is trained directly on the target data with flow classification and length-token retention objectives. The auxiliary head is not required to compute class predictions and can be discarded for deployment.

\begin{itemize}[nosep,leftmargin=*,itemsep=0pt,topsep=2pt]
    \item \Huang{We introduce mixed-resolution packet-length tokenization with a bounded
vocabulary and an auxiliary objective that encourages token-identity
retention in pre-pooling packet features.}
     \item \Huang{We propose a compact end-to-end framework
    integrating packet and
    directional byte encoders with Lorentz prototype classification, trained
    directly on the target data without a separate pretraining stage.
    }

    \item \Huang{Experiments on NUDT-Mobile and DataCon-Website show that Curv-Tail improves Tail-F1 by \textbf{2.91\%} and
    \textbf{1.89\%} over the strongest baselines while using
    \textbf{98.48\%} fewer parameters than MM4Flow with
    \textbf{10.2}$\times$  inference throughput.}
\end{itemize}

\section{Methodology}
\label{sec:method}

Fig.~\ref{fig:curv_tail_architecture} shows the overall architecture of Curv-Tail.
First, each bidirectional flow is represented by mixed-resolution packet-length
tokens and directional byte patches. Second, two compact temporal encoders
produce contextual features; an auxiliary length-token objective is applied
to packet features before pooling. Finally, the pooled representations are
concatenated and projected, and the resulting flow representation is
classified using trainable Lorentz prototypes.

\subsection{Mixed-resolution Packet and Byte Views}
\label{sec:length_encoding}
\Huang{Curv-Tail represents each flow
using a packet view based on discrete length tokens,
direction, and inter-arrival time,
and a byte view based on local payload patterns in the observed bi-directional sequences.
}

\noindent \Huang{\textbf{Packet View:}} For packet $i$, let $\ell_i$  be the
magnitude of its signed size, $d_i$ its direction, and $\Delta t_i$ its
inter-arrival time. We map $\ell_i$ to a mixed-resolution token
$q_i=Q(\ell_i)$:
\begin{equation}
Q(\ell)=
\begin{cases}
\ell, & 0\leq \ell\leq B,\\
B+\operatorname{clip}\!\left(
\operatorname{round}\!\left(\frac{\ell+1}{S}\right)-1,1,G\right),
& \ell>B,
\end{cases}
\label{eq:length_token}
\end{equation}
\Huang{where} $B$ separates the exact and coarse ranges, $S$ controls the coarse
resolution, and $G$ bounds the number of coarse tokens\Huang{. Lengths up to $B$ retain exact identities, while larger lengths
share the $G$ coarse tokens}. Each packet token is combined with embeddings of direction,
log-transformed inter-arrival time, and position.

\noindent \Huang{\textbf{Byte View:}} The upstream and downstream bytes are
embedded separately and compressed into non-overlapping patches by a strided
1-D convolution before positional encoding.

\subsection{Compact Contextualization and Length Retention} \label{sec:context_retention} 
\Huang{The two views are contextualized by two separately parameterized encoders that share the same compact design. } Each block maps its state to the tangent space at the Lorentz origin, applies a gated residual dilated convolution, and maps the result back with the corresponding exponential map.

For fixed width, depth, and kernel size, the temporal encoding cost is linear in sequence length.
After the final logarithmic map, packet features are aggregated by learned-query attention and byte features by masked mean pooling\Huang{, and their concatenation is projected to the fused flow representation $\mathbf h_n$.} 

To encourage pre-pooling retention of length-token identity,
a shared linear head $A_{\mathrm{ret}}$ predicts the observed token
$q_{ni}$ from the contextualized packet feature $\mathbf u_{ni}^p$
at every position before pooling. This objective complements
flow-level supervision, which does not explicitly require these
features to preserve length-token identities before pooling.
The observed token is neither masked nor corrupted; the auxiliary task is an information-retention regularizer, not masked-token reconstruction\Huang{, and it remains class agnostic, neither resampling the data nor changing class priors}.

\subsection{Lorentz Prototype Classification and Objective}
\label{sec:prototype_classification}

\Huang{Classification compares each flow with trainable class
prototypes under the Lorentz model.} 
The fused representation $\mathbf h_n$ and each trainable class vector
$\mathbf r_k$ are mapped by the origin exponential map $\mathcal E_c$ to
$\mathbf z_n$ and prototype $\mathbf p_k$, respectively. 
Their class score is a scaled negative squared Lorentz distance
plus a class bias:
\begin{equation}
\begin{aligned}
\mathbf z_n&=\mathcal E_c(\mathbf h_n),\quad
\mathbf p_k=\mathcal E_c(\mathbf r_k),\\
s_{nk}&=-\frac{1}{\tau c}\operatorname{arcosh}^{2}
\!\left(-c\langle\mathbf z_n,\mathbf p_k\rangle_L\right)+\beta_k,
\end{aligned}
\label{eq:prototype_logits}
\end{equation}
where $\langle\mathbf x,\mathbf y\rangle_L=-x_0y_0+
\sum_{j=1}^{m}x_jy_j$ \Huang{with spatial dimension $m$}, $\tau$ is a
learned positive temperature, and $\beta_k$ is a class bias.

The bounded curvature magnitude $c>0$ (curvature $-c$) is learned jointly and shared by both encoders and the classifier.

For minibatch $\mathcal B$ and its valid packet positions $\mathcal V$, the
end-to-end objective \Huang{can be expressed as:}
\begin{equation}
\mathcal L=\frac{1}{|\mathcal B|}\sum_{n\in\mathcal B}
\operatorname{CE}(\mathbf s_n,y_n)
+\frac{\lambda_{\mathrm{ret}}}{|\mathcal V|}
\sum_{(n,i)\in\mathcal V}
\operatorname{CE}\!\left(A_{\mathrm{ret}}(\mathbf u_{ni}^p),q_{ni}\right),
\label{eq:joint_objective}
\end{equation}
where $\mathbf s_n$ contains the prototype scores, $y_n$ is the flow label,
\Huang{$\operatorname{CE}$ denotes cross-entropy, and
$\lambda_{\mathrm{ret}}$ weights the retention term}.
The first term is ordinary flow-level cross-entropy, while the second encourages
packet-length identity to remain decodable before pooling. The auxiliary term
does not enter the class logits or alter the prediction rule.

\begin{table}[!t]
\centering
\caption{Test results (\%), averaged over seeds 42, 2027, and 3407.
ACC, M-F1, and T-F1 denote accuracy, Macro-F1, and Tail-F1, respectively.
$\dagger$ denotes methods using a separate pretraining stage.}
\label{tab:main}

\vspace{3pt}
\fontsize{9}{11.5}\selectfont
\setlength{\tabcolsep}{3.0pt}
\renewcommand{\arraystretch}{1.10}

\begin{tabular}{@{}l *{6}{r}@{}}
    \toprule

    & \multicolumn{3}{c}{
        \makecell{\textbf{NUDT}\\\textbf{Mobile}~\cite{ZhaoShuang24NudtAPP}}
    }
    & \multicolumn{3}{c}{
        \makecell{\textbf{DataCon}\\\textbf{Website}~\cite{datacon2021website}}
    } \\

    \cmidrule(lr){2-4}
    \cmidrule(lr){5-7}

    \textbf{Method}
    & \textbf{ACC}
    & \textbf{M-F1}
    & \textbf{T-F1}
    & \textbf{ACC}
    & \textbf{M-F1}
    & \textbf{T-F1} \\

    \midrule

    GraphDApp~\cite{Shen2021GraphDapp}
    & 53.11 & 38.23 & 12.93
    & 52.08 & 47.72 & 33.83 \\

    EBSNN~\cite{Xiao22EBSNN}
    & 92.73 & 87.05 & 73.85
    & 2.60 & 1.62 & 0.84 \\

    FastTraffic~\cite{xu2023fasttraffic}
    & 2.85 & 0.02 & 0.00
    & 8.85 & 2.20 & 0.00 \\

    mm-CESNET-v2~\cite{Jan2023mmv2}
    & 88.55 & 79.30 & 58.90
    & 62.87 & 58.34 & 44.41 \\

    \addlinespace[1pt]
    \cmidrule(lr){1-7}
    \addlinespace[1pt]

    ET-BERT$^\dagger$~\cite{lin2022bert}
    & 93.83 & 90.40 & 82.79
    & 2.63 & 2.19 & 1.20 \\

    YaTC$^\dagger$~\cite{zhao2023yet}
    & 90.69 & 81.28 & 56.78
    & 3.40 & 1.86 & 1.32 \\

    TrafficFormer$^\dagger$~\cite{zhou2025trafficformer}
    & 85.16 & 79.54 & 70.88
    & 3.63 & 0.13 & 0.00 \\

    MM4Flow$^\dagger$~\cite{yang2025mm4flow}
    & 96.08 & 91.83 & 80.73
    & 72.31 & 64.33 & 41.10 \\

    \midrule

    \textbf{Curv-Tail (ours)}
    & \textbf{96.70}
    & \textbf{93.48}
    & \textbf{85.70}
    & \textbf{73.53}
    & \textbf{66.66}
    & \textbf{46.30} \\

    \bottomrule
\end{tabular}

\end{table}

\begin{figure}[t]
    \centering
    \label{fig:nudt_class_frequency}
    \scalebox{1.2}[1.2]{%
        \includegraphics[width=0.7\columnwidth, trim=13 0 0 0, clip]{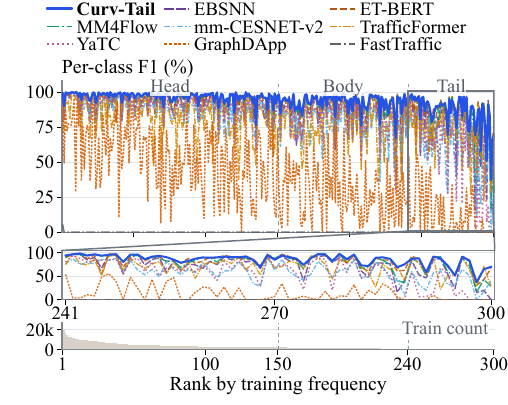}%
    }

    \caption{Per-class test F1 on NUDT-Mobile, ranked by training
frequency. The middle strip zooms into ranks 241--300; the bottom
shows training counts.}
\label{fig:nudt_class_frequency}

\end{figure}

\section{Evaluation}
\label{sec:eval}

\subsection{Experimental Setup}
\label{sec:experimental_setup}

\noindent\textbf{Datasets and baselines.}
We use two naturally long-tailed datasets:
DataCon-Website~\cite{datacon2021website}
(100 website classes over encrypted tunnels) and
NUDT-Mobile~\cite{ZhaoShuang24NudtAPP} (300 mobile-app classes).
After bidirectional five-tuple grouping and removal of flows with
fewer than five packets, they contain 85,146 and 1,282,557 flows,
with training-set imbalance ratios (maximum/minimum class counts)
of 108.92 and 1281.50, respectively.
All methods share a fixed, class-stratified 7:1:2
train/validation/test split at the flow level for closed-set classification.
We compare the eight pretrained and from-scratch baselines in
Table~\ref{tab:main}, using their model-specific input representations.

\noindent\textbf{Implementation.}
Curv-Tail is trained from scratch using up to 64 packets per flow
and 256 bytes per direction. Checkpoints are selected by validation Macro-F1.
Detailed architecture and training configurations are available
in our open-source code\footnote{\label{fn:curvtail-code}
The source code of Curv-Tail is available at
\url{https://github.com/Al3x0r9an/Curv-Tail}.}.

\noindent\textbf{Metrics and runs.}
We report ACC and Macro-F1 over all classes.
Classes ranked by decreasing training frequency form head/body/tail
groups containing 50\%/30\%/20\% of the classes.
Tail-F1 averages per-class F1 over tail classes;
Tail-R is the fraction of correctly classified test flows among
those whose true labels are tail classes.
Table~\ref{tab:main} averages results over seeds 42, 2027, and 3407;
Table~\ref{tab:tail_ablation} uses seed 42.

\subsection{Classification Performance}
\label{sec:classification_performance}

\noindent\textbf{Overall performance.}
Table~\ref{tab:main} shows that Curv-Tail ranks first on both datasets.
On NUDT-Mobile, Curv-Tail exceeds MM4Flow by 0.62 and 1.65 percentage
points in ACC and Macro-F1, respectively, and ET-BERT by
\textbf{2.91} points in Tail-F1.
On DataCon-Website, it exceeds MM4Flow by 1.22 and 2.33 points in
ACC and Macro-F1, and mm-CESNET-v2 by \textbf{1.89} points in Tail-F1.
Against MM4Flow, the strongest pretrained baseline in DataCon Tail-F1,
the gain is 5.20 points. 
These comparisons concern the reported protocols
and do not by themselves identify the causes of differences among baselines.

\noindent\textbf{Across class frequencies.}
Figure~\ref{fig:nudt_class_frequency} shows per-class results on
NUDT-Mobile. Curv-Tail obtains group-wise Macro-F1 scores of 96.98\%,
92.84\%, and 85.70\% for head, body, and tail classes, respectively.
The corresponding gains over MM4Flow are 0.65, 1.12, and
\textbf{4.97} percentage points. Thus, relative to MM4Flow,
the largest group-wise improvement occurs on the tail classes.

\noindent\textbf{Long-tail corrections.} 

On NUDT-Mobile, we test whether classifier-level corrections to
fine-tuned ET-BERT with its backbone frozen can close the gap
to Curv-Tail (Table~\ref{tab:longtail_comparison_etbert_nudt}).
The best evaluated correction, feature mixup, improves Tail-F1
by only 0.08 percentage points; Curv-Tail still outperforms it
by 3.05 and 2.83 points in Macro-F1 and Tail-F1, respectively.

\begin{table}[t]
    \centering
    \caption{Long-tail corrections on NUDT-Mobile (\%), averaged over
three seeds. The ET-BERT encoder is fine-tuned on the target task
and then frozen for the correction stage. T-R denotes Tail-R.}
    \label{tab:longtail_comparison_etbert_nudt}

    \vspace{3pt}
    \fontsize{9}{11.5}\selectfont
    \setlength{\tabcolsep}{2pt}
    \renewcommand{\arraystretch}{1.10}

    \begin{tabular}{@{}lrrrr@{}}
        \toprule

        \textbf{Method}
        & \textbf{ACC}
        & \textbf{M-F1}
        & \textbf{T-R}
        & \textbf{T-F1} \\

        \midrule

        ET-BERT~\cite{lin2022bert}
        & 93.83 & 90.40 & 76.72 & 82.79 \\

        \midrule

        + Logit adjustment~\cite{menon2020long}
        & 93.79 & 90.35 & 77.05 & 82.77 \\

        + Effective-number reweight~\cite{cui2019class}
        & 93.82 & 90.36 & 77.10 & 82.65 \\

        + cRT~\cite{kang2019decoupling}
        & 92.55 & 87.32 & 81.76 & 72.34 \\

        + Feature mixup~\cite{verma2019manifold}
        & 93.84 & 90.43 & 76.89 & 82.87 \\

        \midrule

        \textbf{Curv-Tail (ours)}
        & \textbf{96.70}
        & \textbf{93.48}
        & \textbf{82.26}
        & \textbf{85.70} \\

        \bottomrule
    \end{tabular}
\end{table}

\subsection{Ablation Studies}
\label{sec:ablation_studies}

Table~\ref{tab:tail_ablation} reports ablations on DataCon-Website using seed 42. Discrete length encoding is the most influential tested component: replacing it with continuous inputs lowers Tail-F1 by 11.04 percentage points. Removing length-token retention reduces Tail-F1 by 1.02 points, supporting explicit token-identity supervision beyond input discretization.

The full Lorentz configuration exceeds the Euclidean encoder with a linear classifier by 2.12 points; learning curvature adds 0.28 points over fixed curvature in this run.

\begin{table}[t]
    \centering
    \caption{Tail-F1 ablations on DataCon-Website (\%) using seed 42.
    D, C, and R denote discrete length-token inputs, learnable curvature,
    and length-token retention. 
}
    
    \label{tab:tail_ablation}

    \vspace{2pt}
    \fontsize{9}{11.5}\selectfont
    \setlength{\tabcolsep}{3pt}
    \renewcommand{\arraystretch}{1.00}

    \begin{tabular*}{\columnwidth}
        {@{\extracolsep{\fill}}lcccr@{}}
        \toprule

        \textbf{Configuration}
        & \textbf{D}
        & \textbf{C}
        & \textbf{R}
        & \textbf{Tail-F1} \\

        \midrule

        Continuous encoding
        & $\times$
        & $\checkmark$
        & $\checkmark$
        & 36.19 \\

        Fixed curvature ($c=1$)
        & $\checkmark$
        & $\times$
        & $\checkmark$
        & 46.95 \\

        w/o length retention
        & $\checkmark$
        & $\checkmark$
        & $\times$
        & 46.21 \\

        Euclidean counterpart
        & $\checkmark$
        & $\times$
        & $\checkmark$
        & 45.11 \\

        \midrule

        \textbf{Curv-Tail (full)}
        & $\checkmark$
        & $\checkmark$
        & $\checkmark$
        & 47.23 \\

        \bottomrule
    \end{tabular*}
\end{table}

\begin{figure}[!t]
    \centering
    \captionsetup{skip=2pt,labelsep=period}

    \includegraphics[width=0.9\linewidth]
        {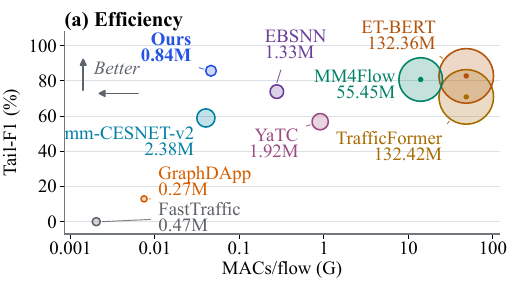}\par
    \nointerlineskip
    \vspace{-1.3mm}
    \includegraphics[width=0.9\linewidth]
        {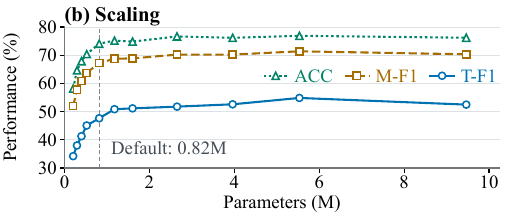}\par

\caption{Efficiency and capacity scaling.
(a) Test Tail-F1 versus counted MACs on NUDT-Mobile;
bubble size indicates parameter count.
(b) Width scaling on DataCon-Website at depth three with seed 42,
reporting validation metrics at the epoch with the best
validation Macro-F1. The dashed line marks the default model.
The default model has 0.84M/0.82M parameters for 300/100 classes,
respectively, with the difference arising solely from
the classification head.}
    \label{fig:efficiency_scaling}
\end{figure}
\subsection{Computational Efficiency and Scaling}
\label{sec:computational_efficiency}

Curv-Tail achieves the highest Tail-F1 among evaluated methods
with 0.84M parameters and 46.07M counted MACs per flow
(Fig.~\ref{fig:efficiency_scaling}(a)).
On an RTX 4090 (FP32, batch 256, 300 classes), its amortized
forward time (median batch time/256) is 0.1359\,ms per flow,
providing 10.2$\times$ and 30.3$\times$ the throughput of
MM4Flow and TrafficFormer, respectively.
Profiling retains the auxiliary head; MAC counts exclude
element-wise operations.

On DataCon-Website (Fig.~\ref{fig:efficiency_scaling}(b)),
scaling from 0.20M to the default 0.82M parameters improves
validation Tail-F1 by 13.44 percentage points.
The 5.53M variant adds 7.20 points at 6.8$\times$ the default
parameter count, while further expansion to 9.46M lowers Tail-F1,
illustrating the compact default's accuracy--capacity trade-off.

\section{Conclusion}
\label{sec:conclusion}
We presented Curv-Tail, combining mixed-resolution length tokens
and pre-pooling retention with compact packet--byte encoders
and Lorentz prototypes. Its three-seed mean Tail-F1 exceeds
the strongest baselines in Table~\ref{tab:main} by 2.91 and
1.89 percentage points on NUDT-Mobile and DataCon-Website.


\linespread{0.95}\selectfont

\bibliographystyle{IEEEbib}
\bibliography{strings,refs}

\end{document}